\documentclass{aastex}
\usepackage{emulateapj5}

\slugcomment{Accepted for publication in ApJ Letters, September
2003.}

\shorttitle{Extrasolar Giant Planets Using FKSI}
\shortauthors{Danchi et al.}

\received{2003 July 16}
\accepted{2003 September 9}
\begin{document}


\title{Detection of Close-In Extrasolar Giant Planets\\
Using the Fourier-Kelvin Stellar Interferometer}


\author{William C. Danchi\altaffilmark{1}, Drake Deming\altaffilmark{2},
Marc J. Kuchner\altaffilmark{3}, Sara Seager\altaffilmark{4}}





\altaffiltext{1}{NASA's Goddard Space Flight Center, Infrared
Astrophysics Branch, Code 685, Greenbelt, MD 20771,
William.C.Danchi@nasa.gov}

\altaffiltext{2}{NASA's Goddard Space Flight Center, Planetary
Systems Branch, Code 693,
ddeming@pop600.gsfc.nasa.gov }

\altaffiltext{3}{ Hubble Fellow,
Department of Astrophysical Sciences, Princeton University,
Princeton, NJ 08544,
mkuchner@astro.princeton.edu }

\altaffiltext{4} { Department of Terrestrial Magnetism,
 Carnegie Institution of Washington,
Washington, DC 20015,
seager@dtm.ciw.edu }


\begin{abstract}

We evaluate the direct detection of extrasolar giant planets with a
two-aperture nulling infrared interferometer, working at angles
${\theta}<{\lambda}/2B$, and using a new `ratio-of-two-wavelengths'
technique.  Simple arguments suggest that interferometric detection
and characterization should be quite possible for planets much closer
than the conventional inner working angle, or angular resolution limit.  We
show that the peak signal from a nulling infrared interferometer of
baseline ($\lesssim 40$ meters) will often occur `inside the null',
and that the signal variations from path-difference fluctuations will
cancel to first order in the ratio of two wavelengths. Using a new
interferometer simulation code, we evaluate the detectability of all
the known extrasolar planets as observed using this two-color method
with the proposed {\it Fourier Kelvin Stellar Interferometer (FKSI)}.
In its minimum configuration {\it FKSI} uses two 0.5-meter apertures
on a 12.5-meter baseline, and a
$\pm 20^{\circ}$ field-of-regard.  We predict that $\sim 7$ known
extrasolar planets are directly detectable using {\it FKSI}, with
low-resolution spectroscopy ($R \sim 20$) being possible in the most
favorable cases.  Spaceborne direct detection of extrasolar giant
planets is possible with $\sim 12$ meter baselines, and does not
require the much longer baselines provided by formation flying.
\end{abstract}



\section{Introduction}

The number of known extrasolar giant planets is now over 100
\citep{marcy03}, almost all of which have been detected
indirectly, using the radial velocity method. Direct detections of
extrasolar giant planets are now becoming possible. For example,
\citet{char02} detected the atmosphere of the `transiting planet'
HD\,209458b using transit spectroscopy \citep{ss00}.  In the case
of planets which do not transit, other techniques of direct
detection are being developed, such as visible coronagraphic
imaging \citep{ks03}, and nulling infrared (IR) interferometry
\citep{bw78, angel97}.  A conventional view is that the
`inner working angle' (IWA) of a nulling interferometer is the
angular distance to the first fringe maximum, i.e. ${\lambda}/2B$,
where $B$ is the interferometer baseline and $\lambda$ is
wavelength. In this {\it Letter}, we show that, when the likely IR
intensities of the extrasolar planets are factored in, nulling IR
interferometers having $B \lesssim 40$ m, are most sensitive to
extrasolar giant planets (EGPs) well inside the conventional IWA,
i.e. `inside the null'.  A substantial number of the known
extrasolar planets are directly detectable using interferometry
with modest baselines, e.g., the $12.5$ m baseline of the minimum
configuration of {\it FKSI} \citep{dan02, danc03}.

\section{Interferometric Detection `Inside the Null'?}


Simple arguments suggest that interferometric detection
can occur for close-in extrasolar giant planets (CEGPs) not fully
resolved spatially by the interferometer. First, it is well known
that a significant source of noise for a spaceborne nulling
interferometer will be photon noise from the leakage of stellar
radiation around the `edges' of the null fringe, due to the finite
angular radius of the star. If leakage from one stellar radius is
a significant noise source, the IR signal from the closest CEGPs,
at tens of stellar radii, will also leak through the null fringe.
Second, we note that ground-based studies of the CEGPs
\citep{rich03} are beginning to achieve the sensitivity needed to
detect the planets in combined light, with no spatial resolution
whatsoever.

For a given interferometer baseline, at what angular separation
does the transmitted IR signal from an extrasolar planet peak? For
decreasing angular separation from the star, the fringe
transmission decreases, but the planet's thermal equilibrium
temperature in the stellar radiation field increases, so the
signal transmitted through the null fringe can remain significant.
Distant planets will be observed with fringe transmission of
unity, but they will be colder. Figure 1 shows the product of the
Planck function at a wavelength of $5\ \mu$m, times the fringe
transmission ($\sin^{2}(\pi \theta B/\lambda)$) for baselines of
8, 12, 20 and 40 m.  The extrasolar system was assumed to be
at 10 pc distance, and the planet was assumed to emit as a
blackbody in thermal equilibrium with a Bond albedo of 0.4.  These
`interferometer contribution functions' were normalized to unity
for the peak signal at the 40-m baseline.  The asterisks on
each curve mark the nominal IWA at $\lambda/2B$.

Figure 1 shows that, even for the longest (40-m) baseline, the
peak signal occurs inside the nominal IWA at $\lambda/2B$.  For
the shorter baselines the peak occurs quite far inside the IWA.
For example, the 12-m peak signal occurs for planets near 0.1 AU,
a factor of 4 below $\lambda/2B$.  For all baselines, the decrease
in signal at the greatest distances occurs because of the lower
planetary temperatures. Although the peak signal for the 12-m
baseline occurs near 0.1 AU, this signal is reduced by about an
order of magnitude from the 40-m case.   Figure 1 suggests that
the sensitivity of a 12-m interferometer working `inside the null'
may be sufficient to allow detection and characterization of some
of the known extrasolar giant planets.

\section{A Two-color Method}


Using a simple analytical model of the nulling interferometer we
demonstrate that the ratio of the intensity at the output of the
nuller at two wavelengths is insensitive to the residual
pathlength fluctuations, provided they are small compared to the
wavelength. A nulling interferometer operates like a conventional
stellar interferometer except that an achromatic $\pi$ phase shift
is applied to the beam coming from one element of the
interferometer and that the beams are symmetrically and
achromatically combined. The equation for the normalized intensity,
$N({\lambda}_1)$, is:

\begin{equation}
N({\lambda}_1) = 1/2 ~ [1 - | V({\lambda}_1) | ~ \cos ~
{\phi}({\lambda}_1)]
\end{equation}

where ${\lambda}_1$ is the first wavelength, $| V({\lambda}_1) |$
is the modulus of the visibility of the source, and ${\phi}(
\lambda _1)= 2 \pi \sigma_N / {\lambda}_1$ is the rms phase error
caused by residual pathlength fluctuations, $\sigma _N$ in the
system. For a point source and no pathlength fluctuations, Eq. 1
reduces to the usual $ \theta ^2$ null of the classical Bracewell
interferometer. A similar equation holds for the output of the
same nuller at a second wavelength, ${\lambda}_2$. Using these
assumptions,

\begin{equation}
| V({\lambda}_1) | \approx 1 - {\pi ^2 \over 16 } \left ( { {
\theta _ {star} } \over { \lambda _1 / B } } \right ) ^ 2
\end{equation}

where ${\theta _ {star} } $ is the angular diameter of the star
and B is the baseline length, and similarly expanding the term for
the phase fluctuations:
\begin{equation}
\cos ~ \phi ( \lambda _1 ) \approx 1 - \phi ( \lambda _1 ) ^2 / 2
= 1 - 2 \pi ^2 \left ( \sigma _N / \lambda _1 \right ) ^2
\end{equation}

Substituting these equations into (1) and using
similar equations for the second wavelength, ${\lambda}_2$, it is
easy to show that:

\begin{equation}
{N( \lambda _1 ) \over N( \lambda _2 )} \approx  \left ( {
{\lambda _2} \over {\lambda _1 } } \right ) ^2
\end{equation}

Hence the stellar leakage and pathlength fluctuations cancel out.

This result is valid for the case when the interferometer is
not rotating about the line of sight or
when the residual pathlength fluctuations occur at much higher
frequencies than the rotation frequency about the line of sight.
The cancellation occurs in the leakage signal variations which
accompany instability in the null fringe (not in the photon
noise), and which would otherwise overwhelm the planetary signal.
This technique is expected to work best when the two wavelengths
$\lambda _1$ and $\lambda _2$ are reasonably close together,
avoiding higher order effects.


\section{An Interferometer Simulation Code}

We have written a simulation code to compute the signal from the
known extrasolar giant planets, as observed by a rotating
2-aperture nulling interferometer, following the principle of
\citet{bw78}, and treating the output signal as a ratio,
exploiting the rationale given above.
The code uses the interferometer parameters
for {\it FKSI} as listed in Table 1.  The modeled fringe pattern
of the interferometer includes path difference errors, based on
the current {\it FKSI} error budget.  Shot-noise is computed from
the total signal, including the stellar leakage, the extrasolar
zodiacal background, and the instrument thermal background. Dark
current and read noise from the detection system are also
included. The stellar spectrum is assumed to be a blackbody, but
wavelength-dependent limb darkening is imposed on the disk, based
on the solar observations of \citet{pierce50}.  The stellar
temperature and radius are estimated from the spectral type.  The
planet is modeled as a blackbody spectrum with superposed
molecular band spectral structure.  The planet's blackbody
equilibrium temperature is computed assuming a Bond albedo of 0.4;
the planetary radius is taken to be 35\% greater than Jupiter,
based on HD\,209458b \citep{brown01}.  The planetary spectral
structure was included by interpolating in the `cloudless'
sequence computed by \citet{sud03}.  The planetary spectrum is
currently assumed to be independent of orbital phase and
inclination to the line of sight.  Orbital motion of the planet is
included, using orbital elements from the Doppler observations,
and a nominal inclination of $45^{\circ}$.  The zodiacal
background is calculated including contributions from
both scattered light and thermal radiation \citep{kuch02}. Since younger
stars will have more massive zodiacal disks, we scaled the mass of
the disk as the -1.76 power of the stellar age \citep{spang01}.
Ages for stars hosting extrasolar planets were taken from
\citet{laws03}. The inclination of the zodiacal disk is also
included (nominally $45^{\circ}$ to the plane of the sky), since
the asymmetry from the inclined disk produces a significant signal
as the interferometer rotates.

\section{Example of an {FKSI} Detection}

We have simulated observations by {\it FKSI} of all known extrasolar
planetary systems
using the code. An example is shown in
Fig. 2, which illustrates the detection of 55 Cancri b.  Figure 2a
shows the number of photons detected versus wavelength, at the {\it
FKSI} spectral resolving power ($\lambda / \delta\lambda = 20$), in
one 300-s integration.  The calculated wavelength range in Fig.
2 extends longward of the nominal wavelength limits of {\it FKSI}.  The
dominant source of photons is stellar leakage, with planetary
radiation second.  The zodiacal radiation falls below the planetary
intensity for this old, and dust-poor system \citep{rayjay02}. The
dominant source of noise depends on wavelength, but is due to stellar
leakage shot noise at the shortest wavelengths, thermal background
radiation noise at the longest wavelength ($8\ \mu$m), and dark current and
read noise at intermediate wavelengths. Since the anticipated total
integration time during an observational `campaign' for a given
planetary system can be many days, planets whose photon counts fall
well below the noise level in a single 300-s integration will
nevertheless be detectable.  But for 55 Cancri b, the planetary signal
photon counts are comparable to the noise photons in a single
300-s integration, for wavelengths $>3\ \mu$m.  This planet is
strongly detectable, and more extensive spectral information could
also be extracted.

In the original Bracewell concept, the rotation of the interferometer
modulates the planetary signal, and we have included this process in
the simulation code.  Figure 2b (lower panel) shows a power spectrum
from a signal time series of 55 Cancri b. As per our two-color method,
the Fourier-transformed signal in this case was the intensity
integrated over wavelengths $6\ \geqslant \lambda\ \geqslant 3\
\mu$m, ratioed to the intensity integrated over $\lambda < 3\ \mu$m.
The denominator in this ratio is dominated by the stellar leakage at
the shortest wavelengths, and contains minimal planetary signal. The
planet peaks at the longer wavelengths in the numerator.  The greater
null depth at long wavelengths enhances this separation of planet and
stellar signals.

The interferometer was rotated slowly, about 15 hours per rotation,
and the simulated campaign lasted about 35 days (50 rotations).  The
zodiacal signal appears in the Fig. 2b power spectrum at twice the
interferometer rotation frequency, i.e. at 37 microHz ($\mu$Hz), and the
planetary signal also contributes at this frequency. Because of the
planet's orbital motion, the planetary signal appears primarily at a
different frequency (see caption), slightly displaced, and {\it
resolved} from the zodiacal signal.  Moreover, because the planet is
inside the null, no overtones are caused by transmission through
higher-order fringes. Note also that, as the signal from the planet
goes in- and out-of-phase with the zodiacal signal, the envelope of
this modulation is seen in the low-frequency region of the power
spectrum as a peak at 1.6 $\mu$Hz, twice the orbital frequency of the
planet.

\section{Discussion}

Our simulations show that a significant number of known
extrasolar planets will be detectable using {\it FKSI}, and that
spectral information can be derived in several of the most favorable
cases.  These simulated detections are robust in the sense that the few most
favorable planetary systems are detectable with almost any reasonable
instrumental configuration (baseline, aperture size, etc.) or choice
of long-wavelength cutoff.  However, detections for the fainter
systems depend significantly on the instrumental parameters and
wavelength range, as well as on the properties of the detectors (dark
current, read noise).

Figure 3 illustrates the distances and orbital semi-major axes for
many of the known planetary systems, with those systems detectable by
{\it FKSI} (using Table 1 parameters) plotted as filled symbols.  The
$\pm 20^{\circ}$ field-of-regard ({\it FOR}) holds seven detectable planets, and
an additional six planets are within a $\pm 40^{\circ}$ {\it FOR}.
The conventional IWA resolutions for baselines of 40, 20, 12,
and 8 m are overplotted.  It can be seen that the detectable
planets are essentially the CEGP systems closest to earth, with
essentially no dependence on the angular resolution limits (IWA
lines).

Our results show that planets can be detected much closer to stars
than expected based on the concept of the inner working angle,
computed from the nominal resolution of the interferometer, i.e.,
$\lambda / 2 B$. This resolution estimate is essentially the
same as the Rayleigh criterion for
conventional telescopes, which is the angular separation
of two stars of equal intensity in which one star is placed at the
first zero of the Airy pattern of the second star.  The Rayleigh
criterion is well known to be a very conservative estimate of
resolution, and sources can be resolved that are substantially
closer than this, using super-resolution techniques such as the
pixon method \citep{pixon}. Similar considerations hold for
interferometers
as CLEAN, MEM, and other methods have provided images with
effective angular resolutions much better than nominally expected \citep{cobb87}.

For a wavelength of 5 $\mu$m, the resolution of the {\it FKSI} is
$ \lambda/2B \approx$ 41 milli-arcsec (mas). But our simulations
demonstrate an `effective resolution' of approximately 1 mas. Our
results are consistent with the working resolution of the
interferometer being determined by the ratio of the rms pathlength
fluctuation to the baseline, $\sigma _N / B$, which for the
parameters in this paper, is 0.2 mas.  This is not as surprising
as it may seem at first, since the pathlength stability requirement to
achieve a $10^{-4}$ null or better is essentially that of an interferometer
with extremely high phase stability.

The work presented here has important implications for the
Terrestrial Planet Finder (TPF) mission \citep{TPFbook}, because
angular resolution is often emphasized over sensitivity. The
desire for a nulling interferometer using free flyer telescopes is
based on the resolution needed to search more than 150 F, G, and K
stars for earth-like planets in the habitable zone
\citep{lunine03}.  Given that the actual IWA can be significantly
smaller than previously thought, means that it may be possible to
achieve the basic goals of TPF with a structurally connected
interferometer having a modest baseline in the range of 20 to 30 m.




\section{Acknowledgements}
We thank
Drs. R. Allen, D. Benford, D. Gezari,
D. Leisawitz, J. Monnier, M. Mumma, L. Mundy, C. Noecker, and W.
Traub,
for their contributions to the {\it FKSI} mission concept.

\clearpage
\begin{table}
\caption{Nominal Parameters for {\it FKSI}}
\vspace{2mm}
\begin{tabular}{ll}
\tableline
Parameter & Value \\
\tableline
Baseline  & 12.5 meters  \\
Aperture  & $2 \times 0.5$ meters diameter  \\
Field of regard & $\pm 20^{\circ}$ \\
Efficiency & 0.05 (electrons out / photons in)  \\
Spectral resolution & $\lambda/\delta\lambda = 20$   \\
Pathlength stability & 15\ nm rms; 10-sec time constant \\
Wavelength range & $1\ \mu$m$\ \leqq\ \lambda\ \leqq\ 6\ \mu$m \\
Optics temperature &  63K     \\
Detector temperature & 35K  \\
Dark current & 0.2 $e^{-}$/sec   \\
Read noise & 8 $e^{-}$  \\
\tableline
\tableline
\end{tabular}
\end{table}

\clearpage

\begin{figure}
\epsscale{0.7}
\plotone{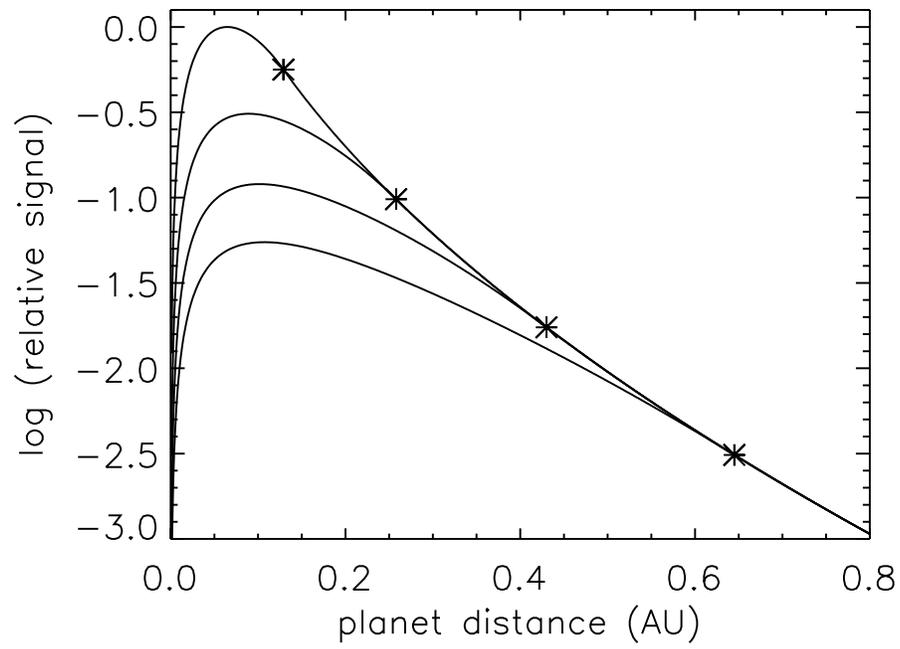}
\vspace{1.0 cm}
\caption{Log of the peak planetary signal seen at $\lambda = 5\ \mu$m by a
nulling interferometer of (upper to lower curves) baseline 40, 20, 12,
and 8 meters. The asterisks mark the nominal inner working angles at
$\lambda/2B$.  The curves were normalized to unity for the peak signal
at the 40-m baseline.  The IR intensity of the planet was
computed from blackbody equilibrium, with Bond albedo =0.4, and
distance 10 pc.\label{fig1}}
\end{figure}

\clearpage

\begin{figure}
\epsscale{0.5}
\plotone{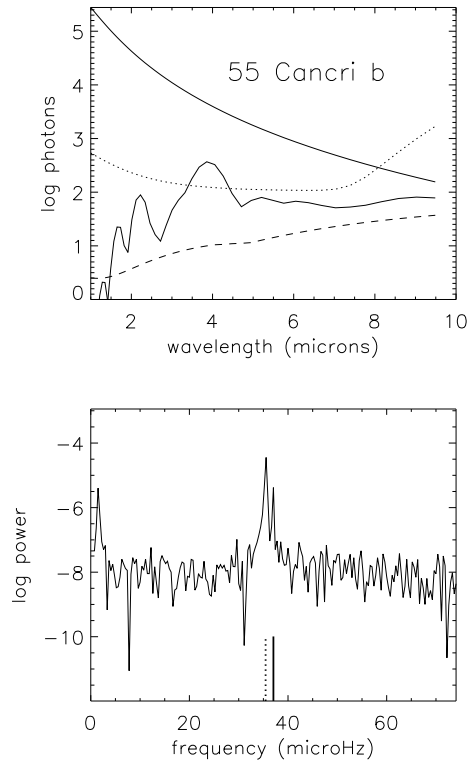}
\vspace{0.5in}
\caption{Simulated observations of the innermost planet in the 55
Cancri system ($a=0.11$ AU, $P=14.65$ days) using {\it FKSI}.  The
upper panel (Figure 2a) shows the number of photons detected at a
single interferometer rotation angle (where the planet signal is
maximum), in a 300 second integration.  The stellar leakage is the
upper solid line, and the detected planetary spectrum is the lower
solid line.  The dashed line is the zodiacal component, and the dotted
line is the total noise.  The lower panel (Figure 2b) shows the power
spectrum from a simulated 35-day observing campaign (50 rotations of
the interferometer).  The marks on the frequency axis indicate
$2\nu_{i}$ (solid mark), and $2|\nu_{i}-\nu_{p}|$ (dashed mark), where
$\nu_{i}$ is the rotation frequency of the interferometer, and
$\nu_{p} = 1/P$, where $P$ is the orbital period of the
planet. \label{fig2}}
\end{figure}

\begin{figure}
\epsscale{0.8}
\plotone{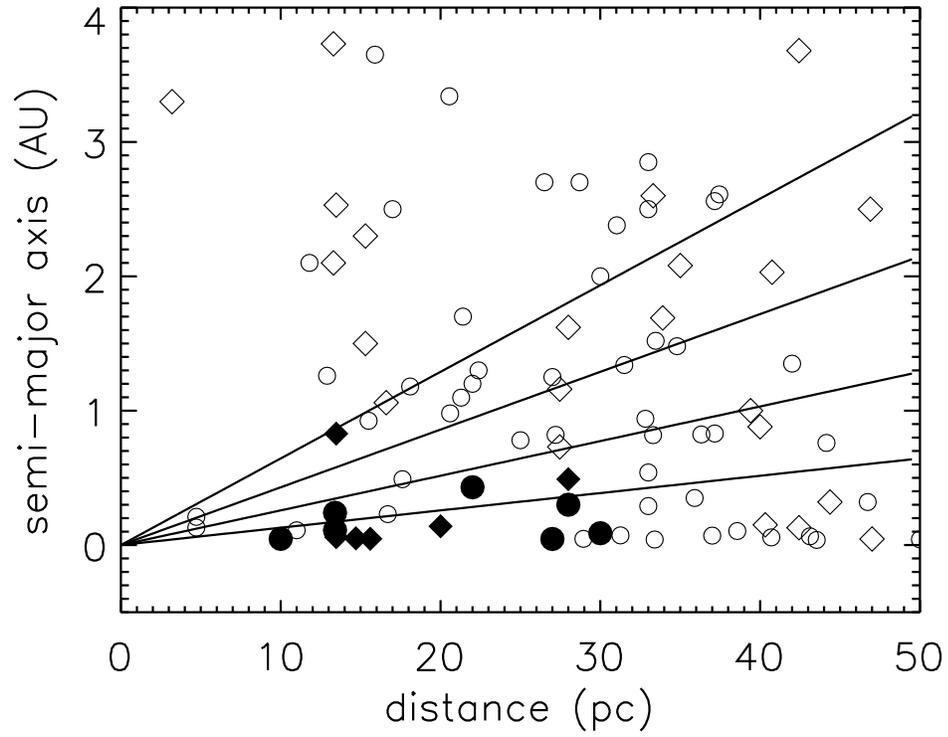}
\caption{Distance (pc) and orbital radius (semi-major axes, in AU) of
known extrasolar planets (having $d \leqslant 50$ pc, and $r \leqslant
4$ AU, and within a $\pm 40^{\circ}$ field of regard).  The planets
nominally detectable by {\it FKSI} are plotted with filled symbols;
circles indicate planets within a $\pm 20^{\circ}$ field of regard,
and diamonds indicate additional detectable planets if the field of
regard is extended to $\pm 40^{\circ}$. The lines correspond to the
first fringe maximum for nulling interferometers of (top to bottom)
baselines 8, 12, 20, and 40 meters, at $\lambda=5\
\mu$m. \label{fig3}}
\end{figure}

\clearpage






\end{document}